# Atomic-Scale Structure Relaxation, Chemistry and Charge Distribution of Dislocation Cores in SrTiO$_3$

*Peng Gao[a,b,c,]\*, Ryo Ishikawa[a], Bin Feng[a], Akihito Kumamoto[a], Naoya Shibata[a], and Yuichi Ikuhara[a,d,e,]\**

[a]Institute of Engineering Innovation, School of Engineering, University of Tokyo, Tokyo 113-8656, Japan.

[b]Electron Microscopy Laboratory, School of Physics, Center for Nanochemistry, Peking University, Beijing 100871, China

[c]Collaborative Innovation Center of Quantum Matter, Beijing 100871, China

[d]Nanostructures Research Laboratory, Japan Fine Ceramic Center, Nagoya 456-8587, Japan

[e]Elements Strategy Initiative for Structural Materials, Kyoto University, Kyoto 606-8501, Japan

\* Corresponding authors. E-mails: p-gao@pku.edu.cn; ikuhara@sigma.t.u-tokyo.ac.jp.

Abstract: By using the state-of-the-art microscopy and spectroscopy in aberration-corrected scanning transmission electron microscopes, we determine the atomic arrangements, occupancy, elemental distribution, and the electronic structures of dislocation cores in the 10° tilted $SrTiO_3$ bicrystal. We identify that there are two different types of oxygen deficient dislocation cores, i.e., the SrO plane terminated $Sr_{0.82}Ti_{0.85}O_{3-x}$ ($Ti^{3.67+}$, 0.48≤x≤0.91) and $TiO_2$ plane terminated $Sr_{0.63}Ti_{0.90}O_{3-y}$ ($Ti^{3.60+}$, 0.57≤y≤1). They have the same Burgers vector of a[100] but different atomic arrangements and chemical properties. Besides the oxygen vacancies, Sr vacancies and rocksalt-like titanium oxide reconstruction are also identified in the dislocation core with $TiO_2$ plane termination. Our atomic-scale study reveals the true atomic structures and chemistry of individual dislocation cores, providing useful insights into understanding the properties of dislocations and grain boundaries.

# 1. Introduction

Dislocations and grain boundaries are ubiquitous in the crystal materials. These defects can have very different atomic arrangement and/or chemistry from the bulk matrix [1-5], which strongly influences on the physical and chemical properties (*e.g.* the ionic and electrical conductivities) or even dominates the entire response of devices that are in nanometer scale. In electroceramic $SrTiO_3$ (STO, a model perovskite oxide), on the basis of the electrical and ionic transport measurements the dislocations are usually assumed to be non-stoichiometric [3] due to the presence of charged defects [6-9]. However, it's still unclear as to what type and amount of charged defect is and how these defects distribute in the dislocations, *i.e.*, whether they are localized in the very core region or spread over the surrounding dislocation area. The atomic arrangements of dislocation cores particularly for the oxygen configuration have been rarely reported [10] mainly due to the experimental limitations and structural complexity of STO and thus a deterministic correlation of the chemical properties to the microstructure for a specific dislocation core has not been achieved.

The knowledge of the local atomic structure and chemistry of the dislocations indeed is extremely difficult to be extracted by the bulk-based characterization techniques such as the electrical measurements [3]. Despite a lot of microscopy efforts [6, 9-15] have also been devoted to reveal the microstructure of dislocations, high-angle annular dark-field (HAADF) [9, 11-13] is insensitive to oxygen, and the conventional TEM [6, 15] and exit surface wave function reconstruction in the negative $C_s$ imaging [10] are unable to distinguish the localized structural reconstruction which commonly exist in the dislocation cores [8, 9, 12, 16]. In contrast, the recent advancements of annular bright-field (ABF) imaging in aberration-corrected scanning transmission electron microscope (STEM) not only enables us to simultaneously visualize both heavier cation and relatively lighter oxygen columns [17] but also is more robust for determining the atomic arrangements in the vicinity of the defects which usually show poor contrast in the HAADF images [18]. In addition, it is also convenient to combine atomic-resolution imaging and spectroscopy such as energy-dispersive X-ray spectroscopy (EDS) and electron energy loss spectroscopy (EELS) in the STEM mode, allowing us to precisely determine the atomic arrangements, occupancy, elemental distribution, and the electronic structures.

Here, by employing these complementary imaging and spectroscopy techniques, we reveal both the cation and anion arrangements in the dislocation cores in STO and identify that the atomic structure of dislocations in STO bicrystal is dominated by the terminated atomic layer on the core, that is, SrO plane terminated core A−$Sr_{0.82}Ti_{0.85}O_{3-x}$ ($Ti^{3.67+}$, 0.48⩽x⩽0.91), and $TiO_2$ plane terminated core B−$Sr_{0.63}Ti_{0.90}O_{3-y}$ ($Ti^{3.60+}$, 0.57⩽y⩽1). Both of them are oxygen deficient and have the same Burgers vector of a[100], while they are distinct in the atomic arrangements and chemical properties. The core B contains high density of Sr vacancies and rocksalt-like reconstruction in the tensile strain zone. Oxygen deficiency in the core A is caused by removal of anions in the confined core zone due to the strong Coulomb repulsive interaction, whereas in the core B the oxygen deficiency mainly originates from the Ti-O polyhedral connection change from the corner sharing in the perovskite to edge sharing octahedrons in the rocksalt-like reconstruction, leading to an increase in the Ti/O ratio (reduced Ti ions). Our study precisely determines the atomic structure and chemistry of non-stoichiometric dislocation cores in $SrTiO_3$. These findings unambiguously clarify a long-standing question on the type and distribution of defects in the dislocations and thus can not only help us to explain the past experiments but also provide essential information for space charge zone calculation and the atomistic simulation for dislocations. The demonstrated methodology by combining the state-of-the-art microscopy and spectroscopy provides unprecedented opportunity to explore the defect properties in complex ceramics.

## 2. Experiments.

*Bi-crystal fabrication*: The STO bi-crystal with a [001]/(100) 10° mistilt grain boundary was fabricated by the thermal diffusion bonding of two STO single crystals. First, 5° off (110) surfaces of the single crystals were polished to a mirror-like state. Then, the surface was cleaned with ethanol and propanol to remove contaminants. Subsequently, one crystal was set on the other to create a 10° tilt grain boundary bi-crystal. Under the uniaxial load of ~0.2 MPa, the two crystals were heat-treated at 700 °C for 20 h at the rate of 20 °C/h in air for bonding. Post annealing was carried out to obtain larger bonded area. Heat treatments were performed at 1000 °C for 80 h in total and subsequently at 700 °C for 14 h.

*TEM sample preparation, images acquisition and analysis*: The TEM specimens were prepared by the mechanical polishing followed by the argon ion milling (Precision Ion Polishing System, Gatan). At the final stage of ion milling, the voltage was set at 0.2 kV for about 5 minutes to remove the surface amorphous layer and minimize the damage. HAADF and ABF images were recorded at 300 kV in a JEM ARM300CF (JEOL Ltd.) with spatial resolutions up to 45 pm. The convergence semi-angle for imaging is 24 mrad, the collection semi-angles snap is 12 to 24 mrad for the ABF imaging and 65 to 240 mrad for the HAADF. During imaging, we also deliberately minimized the electron dose by using small aperture, small beam current and short scanning time. Typical HAADF and ABF images are shown in Figure 1a and b respectively. Two different dislocation cores are visiable, which are labeled as core-A and core-B. To determine the occupancy of individual atomic columns in the dislocation cores, the intensity ratio of each atomic column in the cores to that in the bulk is calculated. We use the relation $I^{1.7}$ for HAADF and $I^{1/3}$ for contrast inverted ABF images to estimate the occupancy, where $I$ is the normalized intensity of the columns in the dislocation cores. For those columns in the dislocation core B with mixed Sr and Ti, both Sr and Ti signal is normalized to the columns in the bulk based on the EDS counts and thus the ratio of Sr to Ti is determined.

*EDS mapping*: The EDS experiments were carried out in a 200 kV JEM-ARM200F (cold-FEG) equipped with dual-SDD EDS detectors (JEOL Ltd.). The convergence semi-angle for imaging is 22 mrad. The total solid angle of EDS detectors is 1.7 sr. Typical net count EDS maps for Sr, Ti and O are shown in Figure 2a-c. No principal component analysis is used to process the data. To estimate the elemental occupancy of the dislocation cores, linear approximation method was employed to calculate the average occupancy for cationic columns. A selected core region of 20×30 pixels is integrated to calculate the average net count. The bulk region is integrated from all over the entire image with the grain boundary region being excluded. The total pixels for bulk calculation are 256×216 pixels. The net count in the cores is normalized to the bulk, *i.e.,* both Sr and Ti are assumed to be 1 in the bulk matrix. The calculated values are 0.82(±0.02) for Sr and 0.85(±0.03) for Ti in the core A compared to the bulk. The error is the standard deviation when averaging the three cores. From two B-type cores, the calculated values are 0.63(±0.04) for Sr and 0.90(±0.01) for Ti compared to the bulk.

*EELS mapping*: The EELS experiments were carried out in a JEM ARM200CF (JEOL Ltd.) equipped with dual Enfinium camera (Gatan). All the spectra were recorded at 200 kV. The electron beam was slightly spread and the acquisition time is 0.1 s/pixel to minimize possible damage to the core structures. The convergence semi-angle is 24 mrad, and the collection angle is 53 mrad. Spectrum image from 400~600 eV with energy dispersion 0.1 eV. The size of mapped region is ~16×16 nm$^2$ with 100×100 pixel. The electron beam is slightly spread to minimize the possible electron beam damage to the dislocation cores. Three 7×5 pixel rectangles are added up as one spectrum to represent the core A. Six 7×5 pixel rectangles are added up as one spectrum to represent the gap between two dislocation cores. Three 7×5 pixel rectangles are added up as one spectrum to represent the core B. The rest regions are used to calculate the spectrum of the grain matrix.

## 3. Results and discussion

### 3.1. ABF image analysis.

Figure 1a is a HAADF (Z-contrast; Z is atomic number) image of a 10° tilted grain boundary in STO bi-crystal that consists of two types of edge dislocations [6, 13, 14, 19]. The core A is SrO plane terminated and core B is TiO$_2$ layer terminated (indicated by the dashed-line arrows in Figure 1a). These two types of dislocations alternately sit along the grain boundary. The distance between two cores is about six unit cells, which is in good agreement with Frank's theory for 10° tilted boundary [20]. Although both core A and B have the same Burgers vector of **a**[100], they show very different contrast, *i.e.*, core B is darker and wider than core A. The ABF image in Figure 1b enables all the columns including oxygen in the dislocation cores to be visible, indicating the ABF is more robust than the HAADF for imaging the defects [18]. Judging from the contrast in the HAADF and ABF images, a substantial decrease in the occupancy occurs in the oxygen column No.11 in the core A in Figure 3a and b, and another oxygen column No.12 at the opposite side of column No.11 is almost unoccupied. The contrast of Sr columns (No. 8, 9, and 10), however, decreases subtly, suggesting the core A is deficient in oxygen. Given the estimated non occupied numbers of 0.45 for the No.11 (O), 1 for the No.12 (O), 0.2 for the No.8 (Sr), 0.09 for the No.9 (Sr), and 0.12 for the No.10 (Sr) columns, we conclude that approximately one oxygen column is missing in the core A (see details in the Experiments section). In the core B in Figure 3c and d, some columns appear diffuse (*e.g.* No.13, No.21, No.22 and No.23) or split (No.16), indicating local structural inhomogeneity. Therefore, to precisely identify the structure of dislocation core B the elemental information is also needed.

**3.2. EDS analysis.**

The atomically resolved EDS maps of the dislocations in Figure 4a-c are averaged from 10 pairs of dislocation cores to minimize the spatial variety (see pristine data in Figure 2a-c). The corresponding net counts profiles are shown in Figure 4d-f. For the core A, the net counts for three elements are lower compared to the bulk matrix. The core B is deficient in Sr and O, whereas the Ti in Figure 2i spreads in the core, confirming the presence of reconstruction that is concentrated in the tensile strain zone. Comparing the Ti map with ABF image in Figure 5, we find that the reconstruction results from a transition of $TiO_6$ octahedrons from corner sharing in perovskite to edge sharing [8, 9, 12, 16], forming a localized rocksalt-like phase.

By quantitatively comparing the average EDS net counts in the core regions with that in the bulk matrix, the average elemental occupancy for the cationic columns can be estimated by linear approximation [21], which gives a reasonable estimation of the local compositions based on the EDS analysis. The Sr is calculated to be ~0.82 for the core A and ~0.63 for the core B, and Ti is ~0.85 for the core A and ~0.90 for the core B, respectively, as shown in Figure 2. Note that the estimated values only represent the average occupancy in core regions compared to the bulk matrix. In fact, the occupancy for each column in the cores can be different in Figure 4a-c. A precise EDS quantification for each atom columns requires spectrum image simulation and first principles calculations, which is very difficult and expensive for such a huge structure unit. Nevertheless, the linear approximation shows that these two dislocation cores have very different compositions. These extracted values are consistent with the line profiles in Figure 4d-e. The under-occupancy of Sr and Ti for both dislocation cores is also consistent with the lower intensity in the Z-contrast image in Figure 1a. Owing to the strong channeling effect and relatively large x-ray absorption, the relative O composition cannot be simply extracted in the case of zone axis, which has been suggested by our previous study [22].

**3.3. EELS analysis.**

Besides the composition, the electrical activities of dislocation cores also strongly depend on the electronic structures of Ti, which can be deduced from the EELS measurement [6, 23]. In Figure 6a, the average spectra from the core A, core B, gap between A and B, and bulk matrix show distinct chemical shift and peak splitting in the Ti L-edges. These features, i.e. the Ti-L edges shift toward the lower energy and less pronounced peak splitting, are indicatives of reduced Ti ions [23]. In the core B, the largest shift and the least peak splitting in the Ti L-edges, and suppressed peaks in the O K-edge in the Figure 6b correspond to the lowest mean valence of Ti. In the gap, shift in the Ti L-edges is also observed because the formation energy of oxygen vacancy in such strained zone is lower than that in the bulk [13, 24, 25].

By fitting the map of Ti L-edge with $Ti^{4+}$ and $Ti^{3+}$ reference spectra in Figure 6c and d, the reduced Ti ions are only found in the grain boundary and mainly concentrated in the dislocation cores. The mean valence of Ti (which is averaged from 7×5-pixel in the core regions equivalent to 1.12×0.8-$nm^2$) in the core A is calculated to be ~3.67, and ~3.60 for the core B. Therefore, combining the EDS and EELS measurements, the dislocation core A is estimated to be $Sr_{0.82}Ti_{0.85}O_{3-x}$ ($Ti^{3.67+}$, $0.48 \leqslant x \leqslant 0.91$) and core B is $Sr_{0.63}Ti_{0.90}O_{3-y}$ ($Ti^{3.60+}$, $0.57 \leqslant y \leqslant 1$). These dislocation cores remain neutral when $x=0.63$ and $y=0.75$.

### 3.4. Discussion

It should be noted that the deduced chemical formulas only represent the average compositions of the dislocation cores. The ABF images in Figure 3 and EDS data in Figure 4 indeed show that each atomic column can have very different occupancy. Generally, the oxygen deficient dislocation cores can be interpreted by the fact that the vacancy formation energy near the dislocation cores is significantly reduced [25]. In fact, the atomic origin of the oxygen deficiency is distinct in these two dislocation cores. For the dislocation core A, a large amount of oxygen is removed from No.11 and No.12 columns because of the strong Coulomb repulsive interaction between these two oxygen columns in the confined core. For the dislocation core B, besides the oxygen deficiency, Sr deficiency and Ti-O polyhedral reconstruction are also observed. The oxygen deficiency in the core B is due to the connection change of the adjacent Ti-O polyhedrons, *i.e.*, from the corner sharing in the perovskite to edge sharing octahedrons in the rocksalt-like reconstruction, increasing the Ti/O ratio (reducing the Ti ions). The formation of such reconstruction originates from the fact that the $TiO_2$ plane terminated core B is rich in Ti and deficient in Sr. Therefore, there is not sufficient Sr to maintain the perovskite-type structure framework and thus the $TiO_6$ octahedrons transform from corner sharing to edge sharing to form rocksalt TiO reconstruction which has larger lattice constant (0.42 nm) compared to the perovskite (0.39 nm) to accommodate the tensile strain [9].

Oxygen deficiency results in reduction of Ti ($Ti^{(4-\delta)+}$ ions) in the dislocation cores, as confirmed by the EELS measurement. Compared to the $Ti^{4+}$ in the $SrTiO_3$ matrix, the reduced $Ti^{(4-\delta)+}$ phase can significantly enhance the electrical conductivity along the dislocation cores, which indeed has been reported by the conducting atomic force microscopy measurements [26]. In addition, the accumulation of oxygen vacancies at the dislocation cores repels the nearby holes or mobile oxygen vacancies [25], creating a depletion zone for oxygen vacancies and holes. As a result, the electronic and ionic conductivities across the dislocations or low angle grain boundary (dislocation arrays) in STO can be affected by the presence of such distribution of space charge. This explains the origins of the back-to-back Schottky barriers at the low angle grain boundary in electroceramic STO [27-29]. Moreover, with the Sr vacancies in STO, Ti-antisite defects could readily form to generate local dipole moments [30-33]. Usually, these dipole moments don't generate macroscopic ferroelectricity as they are buried in an insulating bulk matrix and thus the unscreened depolarization field destabilizes the polarization. However, these dipoles could be aligned by the charges near the non-stoichiometry dislocation cores, generating localized ferroelectricity [34-36], which will affect the electrical activities in a complicated manner.

## 4. Conclusions

In summary, the atomic arrangements for both cations and anions, occupancy, elemental distribution and electronic structures of the <100>{010} dislocation cores in the STO, has been identified by the complementary methods of HAADF, ABF, EDS, and EELS. We find that there are two alternate non-stoichiometric dislocations cores in the low angle STO grain boundary, *i.e.*, core A– $Sr_{0.82}Ti_{0.85}O_{3-x}$ ($Ti^{3.67+}$, $0.48 \leqslant x \leqslant 0.91$) and core B– $Sr_{0.63}Ti_{0.90}O_{3-y}$ ($Ti^{3.60+}$, $0.57 \leqslant y \leqslant 1$), which are distinct in the atomic arrangements and chemistry due to the different terminated atomic layers in the cores. Both of these two dislocation cores are oxygen deficient. For the core A, the oxygen is removed due to the strong repulsive Coulomb interaction between the oxygen columns within the confined core, while in the core B, $TiO_6$ octahedrons in the tensile strain zone transform from the corner sharing to edge sharing to form localized rocksalt-like reconstructions to accommodate the strain. These results reveal the structure relaxation mechanism of the dislocation cores in STO. The type and distribution of charged defects at the dislocation cores and grain boundary are determined. These findings can explain the origins of space charge zone near the dislocation cores, the back-to-back Schottky barrier for low angle grain boundaries, and enhanced electrical conductivities along the dislocation cores [26] in the electroceramic STO, providing critical insights into engineering of dislocations and grain boundaries. The atomic structures of dislocation cores also provide necessary information for future molecular dynamic and *ab initio* simulations. Furthermore, the methodology demonstrated in this work combining complementary advanced microanalysis and microscopy techniques enables precise correlation of the specific microstructure of individual defects with the quantitative chemical properties, providing unprecedented opportunity to explore the defects in complex oxides.

**Acknowledgements**

This work was supported in part by the Grant-in-Aid for Scientific Research on Innovative Areas "Nano Informatics" (Grant No. 25106003), and Scientific Research (A) (15H02290) from Japan Society for the Promotion of Science (JSPS), and Elements Strategy Initiative for Structural Materials (ESISM) and "Nanotechnology Platform" (Project No. 12024046) from the Ministry of Education, Culture, Sports, Science and Technology in Japan (MEXT). P.G. was supported as a Japan Society for the Promotion of Science (JSPS) fellow for part of this work. P.G. is grateful for the support from the National Program for Thousand Young Talents of China and "2011 Program" Peking-Tsinghua-IOP Collaborative Innovation Center of Quantum Matter.

**References**

[1]  Y. Ikuhara. Nanowire design by dislocation technology, Progress in Materials Science 54 (2009) 770-791.

[2]  J.P. Buban, K. Matsunaga, J. Chen, N. Shibata, W.Y. Ching, T. Yamamoto, Y. Ikuhara. Grain boundary strengthening in alumina by rare earth impurities, Science 311 (2006) 212-215.

[3]  K. Szot, W. Speier, G. Bihlmayer, R. Waser. Switching the electrical resistance of individual dislocations in single-crystalline $SrTiO_3$, Nature materials 5 (2006) 312-320.

[4]  N. Shibata, M.F. Chisholm, A. Nakamura, S.J. Pennycook, T. Yamamoto, Y. Ikuhara. Nonstoichiometric dislocation cores in alpha-alumina, Science 316 (2007) 82-85.

[5]  I. Sugiyama, N. Shibata, Z. Wang, S. Kobayashi, T. Yamamoto, Y. Ikuhara. Ferromagnetic dislocations in antiferromagnetic NiO, Nature nanotechnology 8 (2013) 266-270.

[6]  Z. Zhang, W. Sigle, M. Rühle. Atomic and electronic characterization of the a[100] dislocation core in $SrTiO_3$, Physical Review B 66 (2002) 094108.

[7]  S.Y. Choi, J.P. Buban, M. Nishi, H. Kageyama, N. Shibata, T. Yamamoto, S.J.L. Kang, Y. Ikuhara. Dislocation structures of low-angle boundaries in Nb-doped $SrTiO_3$ bicrystals, Journal of Materials Science 41 (2006) 2621-2625.

[8]  J.P. Buban, M. Chi, D.J. Masiel, J.P. Bradley, B. Jiang, H. Stahlberg, N.D. Browning. Structural variability of edge dislocations in a $SrTiO_3$ low-angle 001 tilt grain boundary, Journal of Materials Research 24 (2009) 2191-2199.

[9]  H. Du, C.-L. Jia, L. Houben, V. Metlenko, R.A. De Souza, R. Waser, J. Mayer. Atomic structure and chemistry of dislocation cores at low-angle tilt grain boundary in $SrTiO_3$ bicrystals, Acta Materialia 89 (2015) 344-351.

[10] C.L. Jia, A. Thust, K. Urban. Atomic-Scale Analysis of the Oxygen Configuration at a $SrTiO_3$ Dislocation Core, Physical Review Letters 95 (2005) 225506.

[11] M. Imaeda, T. Mizoguchi, Y. Sato, H.S. Lee, S.D. Findlay, N. Shibata, T. Yamamoto, Y. Ikuhara. Atomic structure, electronic structure, and defect energetics in [001](310)$\Sigma 5$ grain boundaries of $SrTiO_3$ and $BaTiO_3$, Physical Review B 78 (2008) 245320.

[12] K. Takehara, Y. Sato, T. Tohei, N. Shibata, Y. Ikuhara. Titanium enrichment and strontium depletion near edge dislocation in strontium titanate [001]/(110) low-angle tilt grain boundary, Journal of Materials Science 49 (2014) 3962-3969.

[13] S.Y. Choi, S.D. Kim, M. Choi, H.S. Lee, J. Ryu, N. Shibata, T. Mizoguchi, E. Tochigi, T. Yamamoto, S.J. Kang, Y. Ikuhara. Assessment of Strain-Generated Oxygen Vacancies Using $SrTiO_3$ Bicrystals, Nano letters 15 (2015) 4129-4134.

[14] D. Ferre, P. Carrez, P. Cordier. Peierls dislocation modelling in perovskite ($CaTiO_3$): comparison with tausonite ($SrTiO_3$) and $MgSiO_3$ perovskite, Physics and Chemistry of Minerals 36 (2009) 233-239.

[15] Z. Zhang, W. Sigle, W. Kurtz, M. Rühle. Electronic and atomic structure of a dissociated dislocation in $SrTiO_3$, Physical Review B 66 (2002) 214112.

[16] R.F. Klie, W. Walkosz, G. Yang, Y. Zhao. Aberration-corrected Z-contrast imaging of $SrTiO_3$ dislocation cores, Journal of electron microscopy 58 (2009) 185-191.

[17] S.D. Findlay, N. Shibata, H. Sawada, E. Okunishi, Y. Kondo, Y. Ikuhara. Dynamics of annular bright field imaging in scanning transmission electron microscopy, Ultramicroscopy 110 (2010) 903-923.

[18] S.D. Findlay, S. Azuma, N. Shibata, E. Okunishi, Y. Ikuhara. Direct oxygen imaging within a ceramic interface, with some observations upon the dark contrast at the grain boundary, Ultramicroscopy 111 (2011) 285-289.

[19] V. Metlenko, A.H. Ramadan, F. Gunkel, H. Du, H. Schraknepper, S. Hoffmann-Eifert, R. Dittmann, R. Waser, R.A. De Souza. Do dislocations act as atomic autobahns for oxygen in the perovskite oxide $SrTiO_3$?, Nanoscale 6 (2014) 12864-12876.

[20] F.C. Frank. Crystal dislocations-elementary concepts and definitions Philosophical Magazine 42 (1951) 809-819.

[21] Z. Chen, D.J. Taplin, M. Weyland, L.J. Allen, S.D. Findlay. Composition measurement in substitutionally disordered materials by atomic resolution energy dispersive X-ray spectroscopy in scanning transmission electron microscopy, Ultramicroscopy (2016).

[22] N.R. Lugg, G. Kothleitner, N. Shibata, Y. Ikuhara. On the quantitativeness of EDS STEM, Ultramicroscopy 151 (2015) 150-159.

[23] D.A. Muller, N. Nakagawa, A. Ohtomo, J.L. Grazul, H.Y. Hwang. Atomic-scale imaging of nanoengineered oxygen vacancy profiles in $SrTiO_3$, Nature 430 (2004) 657-661.

[24] H.S. Lee, T. Mizoguchi, J. Mistui, T. Yamamoto, S.J.L. Kang, Y. Ikuhara. Defect energetics in SrTiO$_3$ symmetric tilt grain boundaries, Physical Review B 83 (2011) 104110.

[25] D. Marrocchelli, L. Sun, B. Yildiz. Dislocations in SrTiO3: easy to reduce but not so fast for oxygen transport, Journal of the American Chemical Society 137 (2015) 4735-4748.

[26] K. Szot, G. Bihlmayer, W. Speier. Nature of the Resistive Switching Phenomena in TiO$_2$ and SrTiO$_3$: Origin of the Reversible Insulator-Metal Transition. in: Camley RE, Stamps RL, (Eds.). Solid State Physics, Vol 65, vol. 65. Elsevier Academic Press Inc, San Diego, 2014. pp. 353-559.

[27] Y.M. Chiang, T. Takagi. Grain-boundary chemistry of barium-titanate and strontium titanate. 1. hihg-temperature equilibrium space-charge, Journal of the American Ceramic Society 73 (1990) 3278-3285.

[28] R. Waser. Electronic-properties of grain-boundaries in SrTiO$_3$ and BaTiO$_3$ ceramics, Solid State Ionics 75 (1995) 89-99.

[29] M. Kim, G. Duscher, N.D. Browning, K. Sohlberg, S.T. Pantelides, S.J. Pennycook. Nonstoichiometry and the electrical activity of grain boundaries in SrTiO$_3$, Physical Review Letters 86 (2001) 4056-4059.

[30] D. Lee, H. Lu, Y. Gu, S.Y. Choi, S.D. Li, S. Ryu, T.R. Paudel, K. Song, E. Mikheev, S. Lee, S. Stemmer, D.A. Tenne, S.H. Oh, E.Y. Tsymbal, X. Wu, L.Q. Chen, A. Gruverman, C.B. Eom. Emergence of room-temperature ferroelectricity at reduced dimensions, Science 349 (2015) 1314-1317.

[31] G. Burns, F.H. Dacol. Crystalline ferroelectrics with glassy polarization behavior, Physical Review B 28 (1983) 2527-2530.

[32] L.E. Cross. RELAXOR FERROELECTRICS, Ferroelectrics 76 (1987) 241-267.

[33] M. Choi, F. Oba, I. Tanaka. Role of Ti Antisitelike Defects in SrTiO$_3$, Physical Review Letters 103 (2009) 4.

[34] N. Bickel, G. Schmidt, K. Heinz, K. Muller. Ferroelectric relation of the SrTiO$_3$ (100) surface Physical Review Letters 62 (1989) 2009-2011.

[35] V. Ravikumar, D. Wolf, V.P. Dravid. Ferroelectric monolayer reconstruction of the SrTiO$_3$ (100) surface Physical Review Letters 74 (1995) 960-963.

[36] R. Herger, P.R. Willmott, O. Bunk, C.M. Schlepuetz, B.D. Patterson, B. Delley. Surface of strontium titanate, Physical Review Letters 98 (2007) 076102.

FIGURES

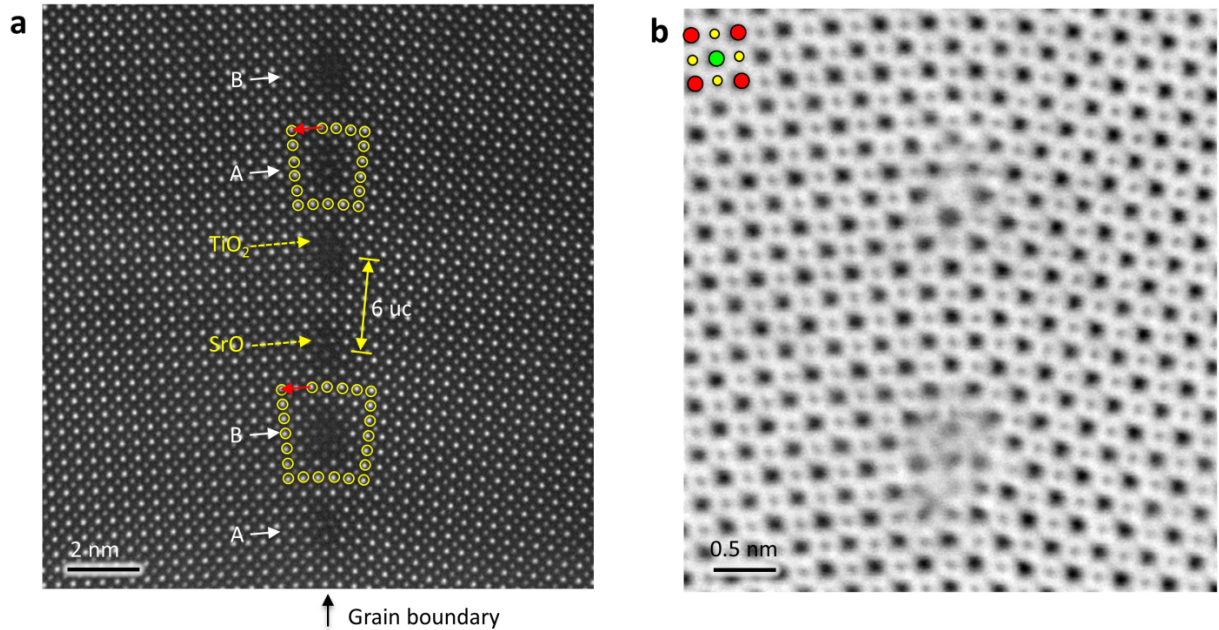

**Figure 1.** A 10° grain boundary in SrTiO$_3$ bi-crystal consisting of dislocation cores. (a) High angle annular dark field (HAADF) image showing two types of dislocation cores in the grain boundary. The distance between these two cores is about six unit cells. The arrows indicate the core A is SrO plane terminated and the core B is TiO$_2$ plane terminated. (b) A higher magnification annular bright field (ABF) image of the same grain boundary.

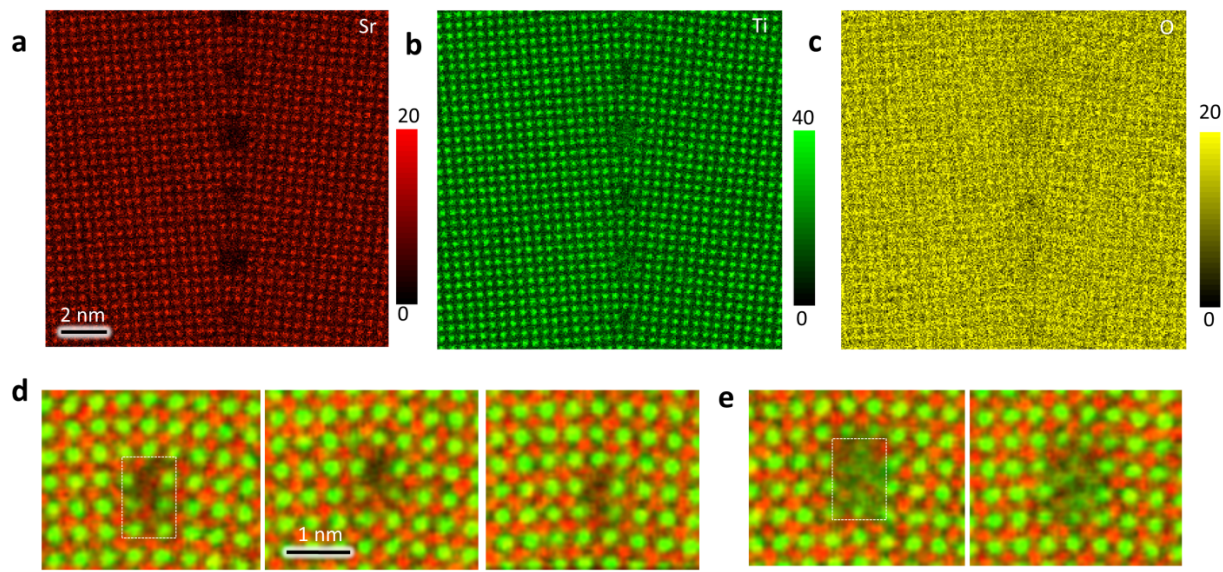

**Figure 2.** Single EDS spectrum of the 10° grain boundary in SrTiO$_3$ bi-crystal. (a) Net count map of Sr. (b) Net count map of Ti. (c) Net count map of O. (d) Three A-type cores. The calculated values are 0.82(±0.02) for Sr and 0.85(±0.03) for Ti in the core A compared to the bulk. The error is the standard deviation (s.d.) when averaging the three cores. (e) From two B-type cores, the calculated values are 0.63(±0.04) for Sr and 0.90(±0.01) for Ti compared to the bulk. The error is the s.d. when averaging the two cores. Note that these estimated numbers only represent the average occupancy in this core region compared to the bulk matrix.

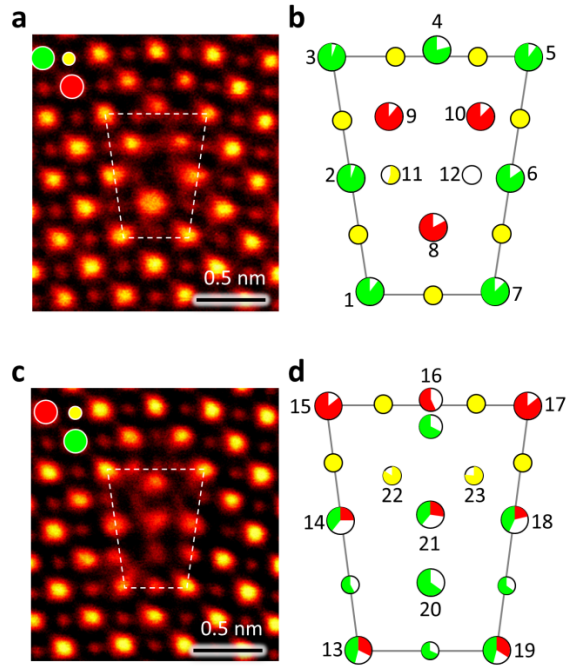

**Figure 3.** Atomic arrangements of dislocation cores in SrTiO$_3$. (a) Enlarged-view of the type A dislocation core. The contrast in the ABF image is inverted for clarity. (b) The corresponding schematic illustration. The occupancy is estimated on the basis of the contrast in the HAADF and ABF images (See Experiments section). (c) Enlarged-view of the type B dislocation core. (d) The corresponding schematic illustration. The occupancy is estimated from the combination of the HAADF, ABF and EDS mapping.

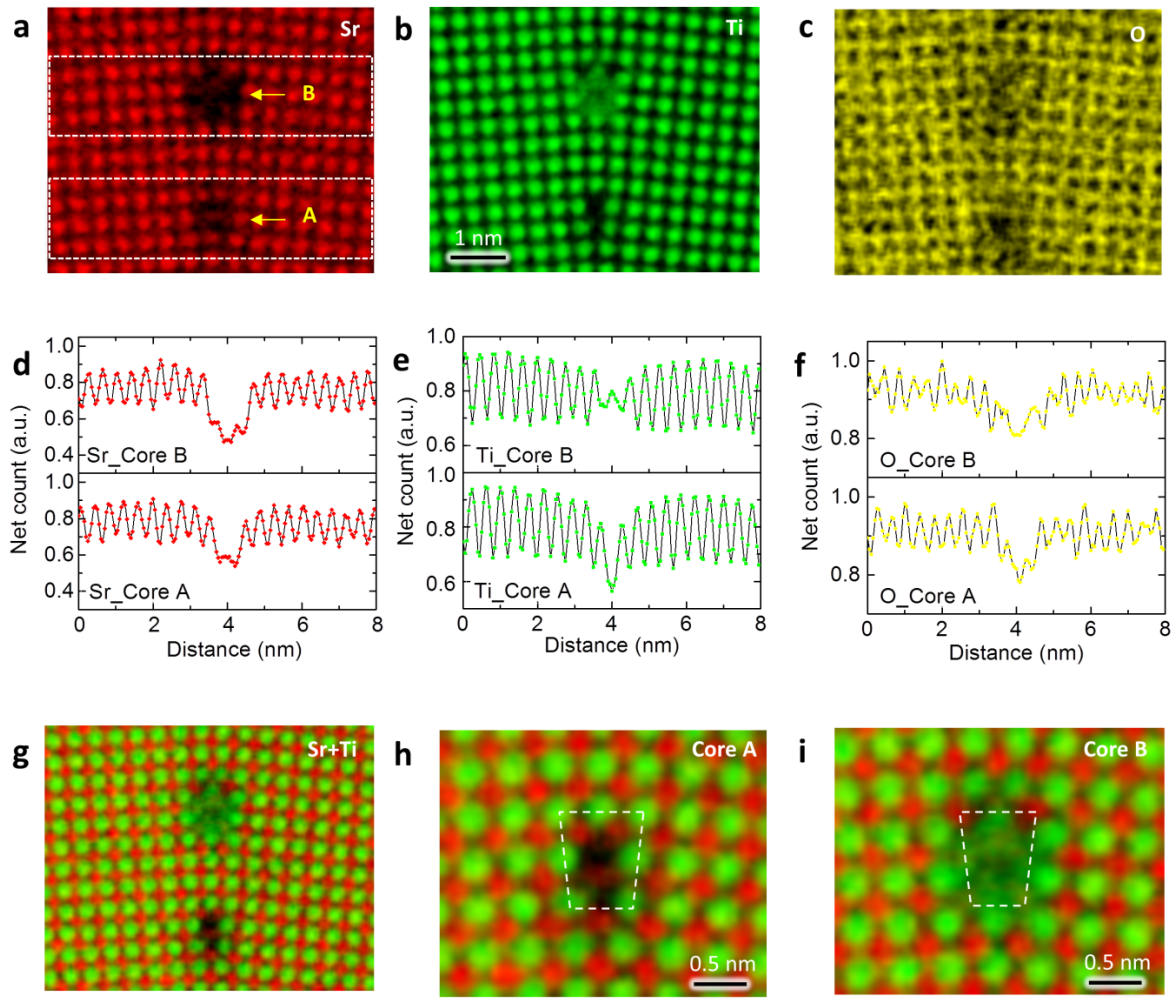

**Figure 4.** Energy dispersive X-ray spectra (EDS) showing the elemental distribution at the dislocation cores in a 10° grain boundary in SrTiO$_3$ bi-crystal. (a) Net count map of Sr. (b) Net count map of Ti. (c) Net count map of O. These maps are added up from 10 pairs of dislocation cores to enhance the signal and minimize the effect of spatial fluctuation. (d) Line profiles of Sr across the dislocation cores. The profiles are integrated from the rectangle regions in (a). (e) Line profiles of Ti across the dislocation cores. (f) Line profiles of O across the dislocation cores. (g) Color mix of Sr (red) and Ti (green). (h) Enlarged view of core A. (i) Enlarged view of core B.

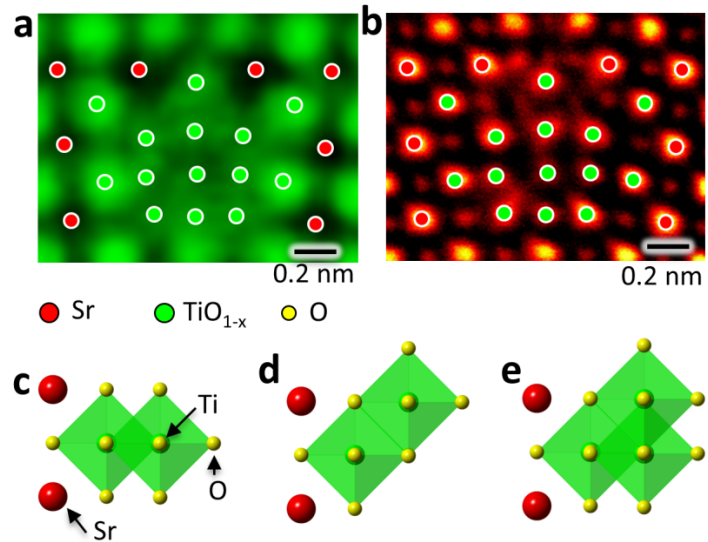

**Figure 5.** Structural reconstruction in the dislocation core B. (a) A structural model overlaid with the Ti net count EDS map. (b) The structural model overlaid with the ABF image. The contrast is inverted for clarity. (c) TiO$_6$ octahedrons transform from corner sharing into edge sharing configuration to form rocksalt-like TiO structure. Such configuration explains the Ti appears in the O columns. (d) Another configuration of edge sharing explains the Ti appears in the Sr columns. (e) A configuration of edge sharing matches the EDS map and ABF image.

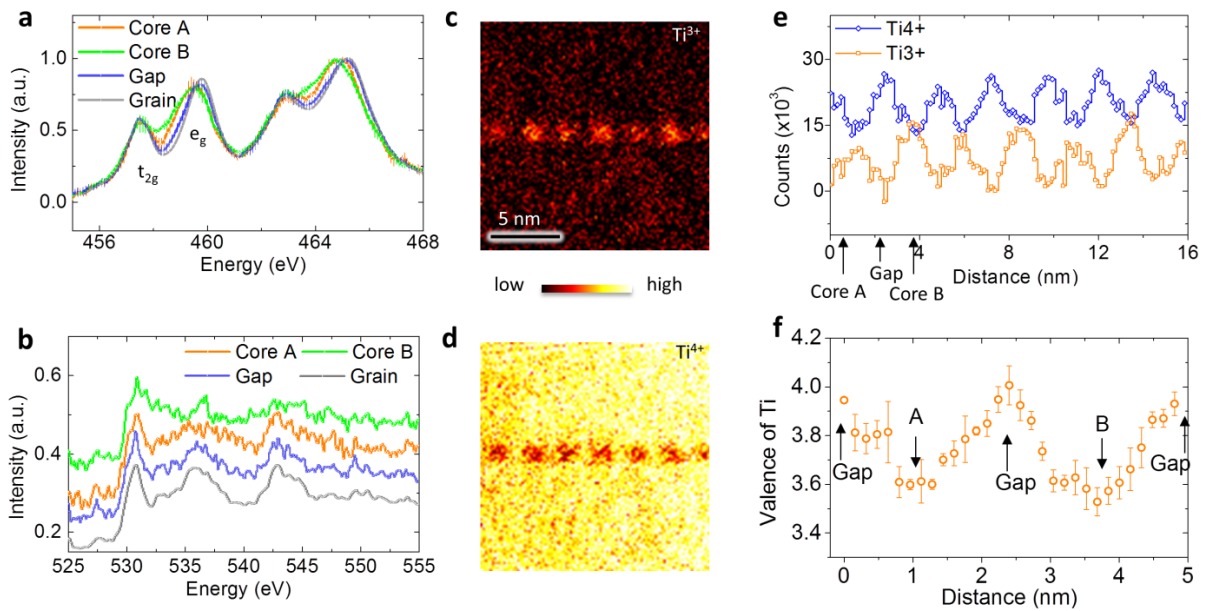

**Figure 6.** Electronic structures of dislocation cores in a 10° grain boundary in SrTiO$_3$ bi-crystal. (a) Average Ti-L edges from three cores A (orange), three cores B (green), six gaps (blue), and two grains (grey). (b) Average O-K edge. The fitted maps of (c) Ti$^{3+}$ and (d) Ti$^{4+}$. (e) The plot of Ti valence shows the distribution of Ti$^{3+}$ and Ti$^{4+}$ along the grain boundary. (f) Average valence distribution across two dislocation cores along the grain boundary. The mean valence of Ti in the core A is calculated to be ~3.67, and ~3.60 for the core B.